%% file: rffrmjxydxptpxfrdpdkcjnfmqmgdtdf/main.tex
\documentclass[10pt,twocolumn,letterpaper]{article}

\usepackage[pagenumbers]{cvpr} 
\usepackage{cvpr}

\usepackage{capt-of}
\usepackage[dvipsnames]{xcolor}
\usepackage{xspace}
\usepackage{enumitem}
\usepackage{amsmath,amssymb,amsbsy,amsfonts,dsfont,pifont,bm,bbm,mathrsfs,mathtools,nicefrac}
\usepackage{booktabs,multirow,adjustbox,diagbox,threeparttable,tabularray}
\usepackage[most]{tcolorbox}

\newtcolorbox{promptbox}{
  colback=gray!5,
  colframe=gray!40,
  boxrule=0.4pt,
  arc=2pt,
  left=4pt,
  right=4pt,
  top=4pt,
  bottom=4pt,
}

\newcommand{\tocite}[1]{{\color{red} [TO CITE]}}

\newcommand{\supp}{\textit{Supplementary Material}\xspace}
\usepackage[ruled]{algorithm2e} %

\usepackage[table]{xcolor}
\newcommand{\methodname}{SURF}
\newcommand{\method}{\methodname\xspace}
\newcommand{\priorname}{signatures}
\newcommand{\prior}{\priorname\xspace}
\usepackage{multirow}
\usepackage{makecell}
\usepackage{graphicx}
\usepackage{subcaption}
\usepackage{caption}    %
\usepackage[absolute,overlay]{textpos} %
\usepackage{xspace}
\usepackage{bm}
\usepackage{marvosym} 

\definecolor{cvprblue}{rgb}{0.21,0.49,0.74}
\usepackage[pagebackref,breaklinks,colorlinks,citecolor=cvprblue]{hyperref}
\usepackage{wrapfig}
\usepackage[capitalize]{cleveref}  
\crefname{section}{Sec.}{Secs.}
\Crefname{section}{Section}{Sections}
\crefname{table}{Tab.}{Tabs.}
\Crefname{table}{Table}{Tables}
\crefname{figure}{Fig.}{Figs.}
\Crefname{figure}{Figure}{Figures}
\crefname{equation}{Eq.}{Eqs.}
\Crefname{equation}{Equation}{Equations}
\def\paperID{1712}
\def\confName{CVPR}
\def\confYear{2026}

\title{
\includegraphics[height=6mm]{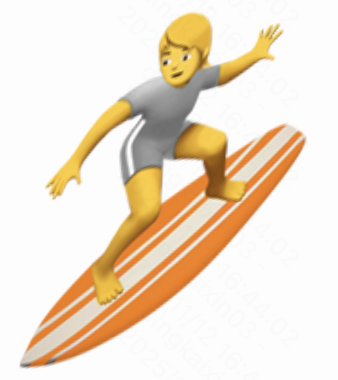}\  SURF: \underline{S}ignat\underline{u}re-\underline{R}etained \underline{F}ast Video Generation}
\author{%
Kaixin Ding\textsuperscript{$\rm 1^{\dagger}$}, 
Xi Chen\textsuperscript{\rm 1}, 
Sihui Ji\textsuperscript{\rm 1}, 
Yuan Gao\textsuperscript{\rm 2}, 
Liang Hou\textsuperscript{\rm 2}, \\
Xin Tao\textsuperscript{\rm 2}, 
Hengshuang Zhao\textsuperscript{\rm 1 \Letter} \\
\textsuperscript{\rm 1}The University of Hong Kong,~\textsuperscript{\rm 2}Kling Team, KuaishouTechnology
}
\begin{document}

\twocolumn[{
\renewcommand\twocolumn[1][]{#1}
\maketitle
\begin{center}
    \vspace{-15pt}
    \includegraphics[width=1.0\linewidth]{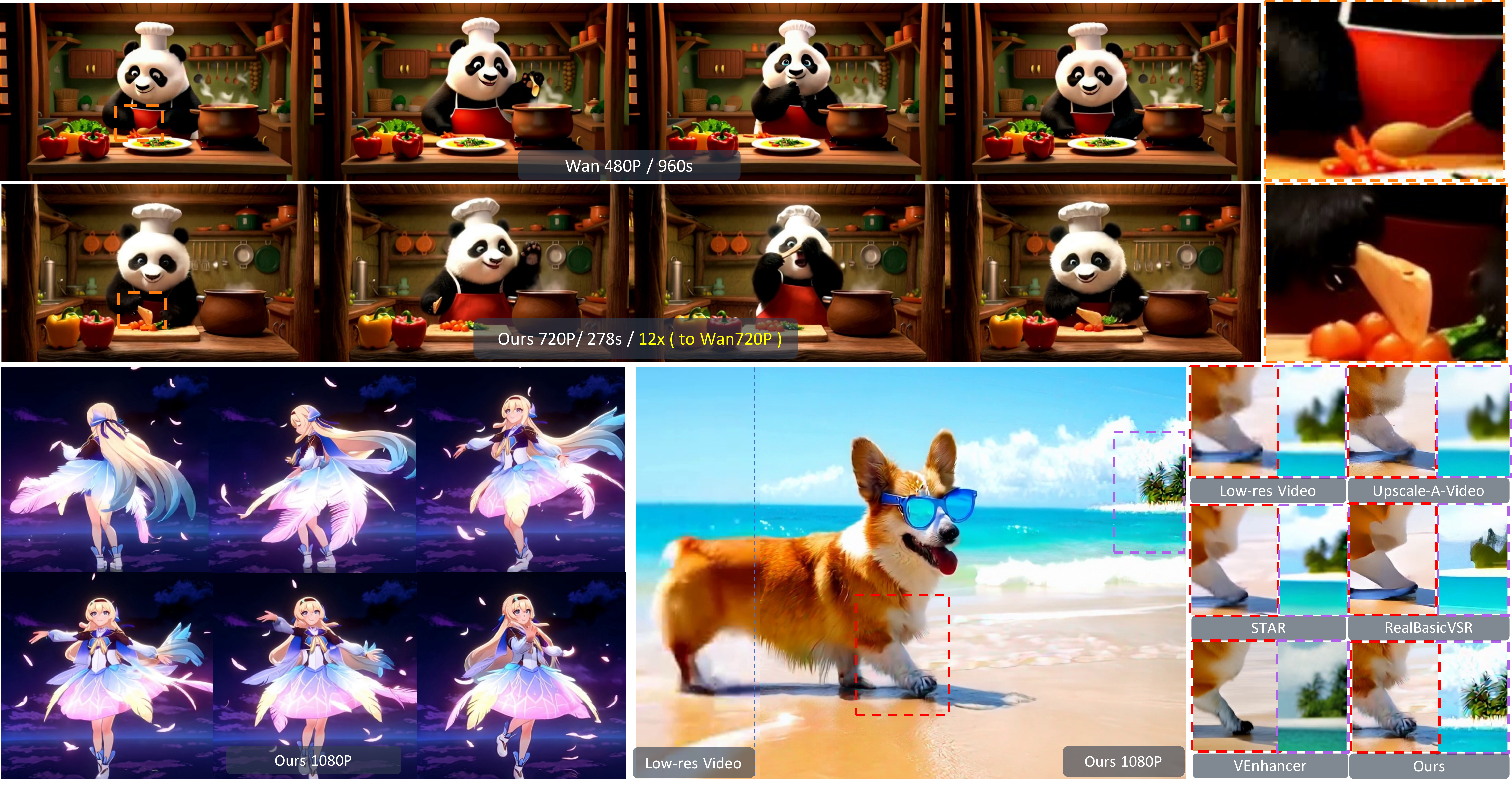}
    \captionsetup{type=figure}
    \vspace{-20pt}
    \caption{%
    \textbf{\method generates high-resolution and high-quality videos efficiently}. 
    It acts as a simple plug-in and retains the \prior~(\textit{i.e.,} layout, semantic, motion, \textit{etc.}) of the baseline model with more than 12× speed-up. 
    More results are available on \href{https://kxding.github.io/project/SURF/}{\textbf{Project Page}}. 
    }
    \label{fig_1}
    \vspace{-6pt}
\end{center}
}]

\let\thefootnote\relax
\footnotetext{\noindent\hspace{-1em}$\dagger$ Work done at Kling Team, Kuaishou Technology.}
\footnotetext{\noindent\hspace{-1em}\Letter~Corresponding Author}

\setlength{\abovedisplayskip}{3pt}
\setlength{\belowdisplayskip}{3pt}
\input{sections/0_abstract}
\input{sections/1_introduction}
\input{sections/2_related_works}
\input{sections/3_method}
\input{sections/4_exp}

\input{sections/5_conclusion} 
\clearpage
\noindent \textbf{Acknowledgements.}
This work is supported by the National Natural Science Foundation of China (No. 62422606, 624B2124) and Hong Kong Research Grant Council General Research Fund (No. 17213925).

\bibliography{main}
\bibliographystyle{conference}

\end{document}

%% file: sections/0_abstract.tex
\vspace{-10pt}
\begin{abstract}
\vspace{-10pt}
\noindent The demand for high-resolution video generation is growing rapidly. 
However, the generation resolution is severely constrained by slow inference speeds. For instance, Wan~2.1 requires over 50 minutes to generate a single 720p video.
While previous works explore accelerating video generation from various aspects, most of them compromise the distinctive \prior~(\textit{e.g.,} layout, semantic, motion) of the original model.
In this work, we propose \method, an efficient framework for generating high-resolution videos, while maximally keeping the \prior. 
Specifically, \method divides video generation into two stages: 
First, we leverage the pretrained model to infer at optimal resolution and downsample latent to generate low-resolution previews in fast speed; then we design a Refiner to upscale the preview.
In the preview stage, 
we identify that directly inferring a model~(trained with higher resolution) on lower resolution causes severe losses in signatures.
So we introduce noise reshifting, a training-free technique that mitigates this issue by conducting initial denoising steps on the original resolution and switching to low resolution in later steps.
In the refine stage, we establish a mapping relationship between the preview and the high-resolution target, which significantly reduces the denoising steps. We further integrate shifting windows and carefully design the training paradigm to get a powerful and efficient Refiner. 
In this way, \method enables generating high-resolution videos efficiently while maximally closer to the \prior of the given pretrained model.
\method is conceptually simple and could serve as a plug-in that is compatible with various base model and acceleration methods.
For example, it achieves 12.5× speedup for generating 5-second, 16fps, 720p Wan 2.1 videos and 8.7× speedup for generating 5-second, 24fps, 720p HunyuanVideo.
\vspace{-15pt}
\end{abstract}

%% file: sections/1_introduction.tex
\begin{figure*}[t]
\centering 
\includegraphics[width=0.98\linewidth]{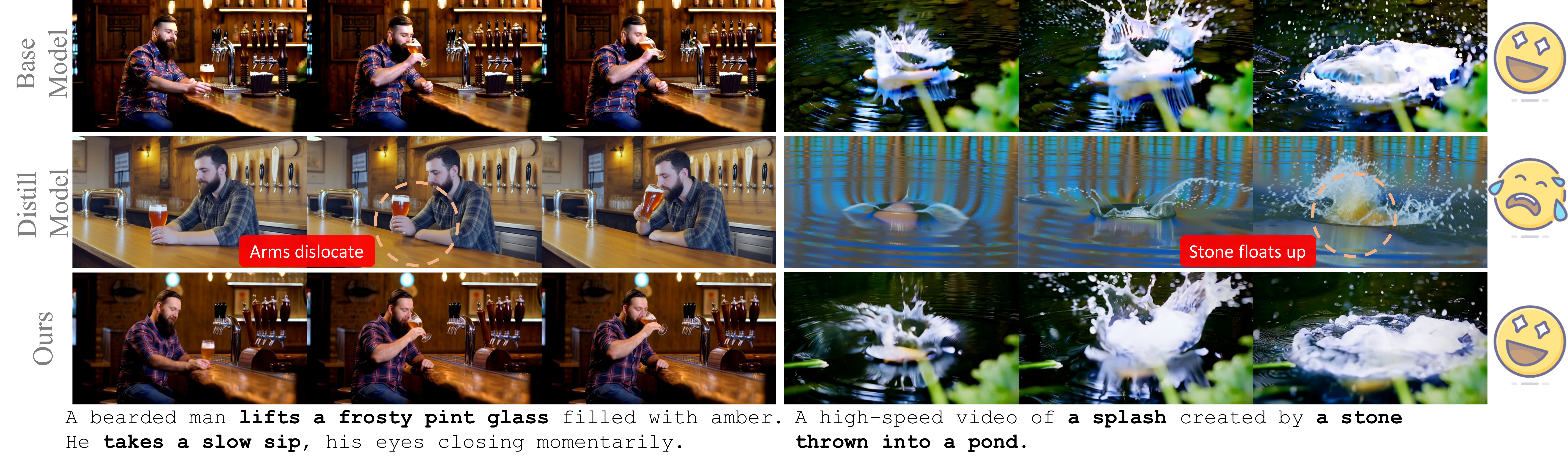} 
\vspace{-10pt}
\caption{
    \textbf{Demonstration for signature retention.} Distilled model loses the signatures of the base model, which could lead to misaligned body parts and weak semantic consistency. Our method maintains the layout, semantics, and motion of base model with huge speed up. 
}
\label{fig:optimal_res}
\end{figure*}

\section{Introduction}\label{sec:intro}
Video generation~\citep{yang2024cogvideox,kuaishou2024klingai,wan2025} has witnessed remarkable advancements in recent years, with sophisticated models continually pushing the boundaries of quality and capability. 
However, the computational cost remains a critical bottleneck, significantly hindering further advances towards higher resolution and richer details.  
For example, it costs roughly 50~min to generate a 5s 720p video for the current SoTA video generation models~\citep{wan2025, kong2024hunyuanvideo} and still 6~min for distillation model~\citep{zhang2025accvideo}. %

Facing this challenge, existing researches explored various strategies to make the generation process more efficient.
Specifically, methods~\citep{luo2023latent,starodubcev2025scale} leverage step distillation to reduce the total denoising steps. 
Studies like \citep{zhang2025spargeattn,ding2025efficientvditefficientvideodiffusion} primarily focus on attention sparsity to improve efficiency. 
Furthermore, cascade multi-scale generation techniques \citep{ren2024ultrapixel,chen2025pixelflow} have been proposed to enhance the efficiency of high-resolution image generation. 

Although these methods could improve the generation efficiency, most of them inevitably compromise the intrinsic \prior of the original model as shown in Fig.~\ref{fig:optimal_res}.    
The model's characteristics are reflected in its preferences for aesthetic style, semantically aligned layout and motion dynamics, \textit{etc.} 
Preserving these characteristics are significant when accelerating a given model, as they act as a prior of the model and could directly reflect the generation quality.   

To achieve this goal, we propose an efficient framework for high-resolution video generation, called \method. This framework divides the whole process into two stages based on their natural features. 
In \textbf{Preview stage}, we leverage a powerful pretrained model to infer at optimal resolution and latent downsample to generate low-resolution previews. The pre-k steps focus on high noise period which determines the main content of the entire video and the post-k steps uses the low resolution latent to gain speed.
We analyze that each pretrained model has its own optimal resolution~(usually the training resolution).  
While distilling the model could increase speed in a large, it often lacks access to the large-scale pretraining data and thus fails to preserve crucial \prior.
Inspired by the observation that the early denoising steps determine the overall content, and the later steps refine the details, we introduce noise reshift method to address this problem.   
Specifically, we start from the pretrained model's optimal resolution in the early denoising steps, then switch to a lower resolution for the remaining steps.
In this way, the low-resolution preview could be efficiently generated and preserve the \prior of the pretrained model.
In \textbf{Refine stage}, we train a powerful and efficient Refiner through establishing a mapping relationship between the low-resolution preview and the high-resolution target.
It significantly reduces the NFE~(number-of-evaluations) to 10 while enriching details and correcting unreasonable artifacts.
In addition, we also integrate the shift window method into the upscaling model to further reduce the computation.

As shown in Fig.~\ref{fig_1}, our method generates high-detailed and high-quality videos with signatures~(layout, semantic, motion, \textit{etc}.) closer to the base model (Wan~2.1) and achieves a huge speedup.
Besides, our \method paradigm allows users to generate multiple previews efficiently at the same time and pick the satisfied content for further refinement to arbitrary size.
Experiments show that \method achieves 12.5× speedup compared with Wan~2.1 14B for generating 5-second, 16fps, 720p videos and 8.7× speedup compared with HunyuanVideo for generating 5-second, 24fps, 720p videos.

%% file: sections/2_related_works.tex
\section{Related Work}\label{sec:related}

\noindent \textbf{High-resolution image and video generation.}  
Existing approaches can be broadly categorized into training-based and training-free paradigms. 
Training-based methods combine architectural innovations and model fine-tuning strategies. 
Turbo2K~\citep{ren2025turbo2kultraefficienthighquality2k} accelerates 2K video synthesis by leveraging a compressed latent space and knowledge distillation within a hierarchical two-stage framework, ensuring structural coherence.
UltraPixel~\citep{ren2024ultrapixeladvancingultrahighresolutionimage} generates 4K-resolution images using cascade diffusion models that feature shared parameters and scale-aware layers, minimizing additional parameters for high-resolution outputs. 
PixArt-$\Sigma$~\citep{chen2024pixartsigmaweaktostrongtrainingdiffusion} and ResAdapter~\citep{cheng2025resadapter} enhance base models through fine-tuning but remain resolution-constrained.
Other approaches such as ResMaster~\citep{shi2024resmastermasteringhighresolutionimage} and HiPrompt~\citep{liu2024hiprompttuningfreehigherresolutiongeneration} introduce multi-modal prompting mechanisms at the expense of computational efficiency.
Training-free methods adapt inference strategies or architectures without retraining. 
MultiDiffusion~\citep{bartal2023multidiffusionfusingdiffusionpaths} and its variants (\textit{e.g.}, SyncDiffusion~\citep{lee2023syncdiffusion}, Demofusion~\citep{du2024demofusion}) employ sliding-window denoising but suffer from repetition or computational redundancy.
ScaleCrafter~\citep{he2023scalecrafter}, recitifiedHr~\citep{yang2025efficient}, FouriScale~\citep{huang2024fouriscalefrequencyperspectivetrainingfree}, and HiDiffusion~\citep{zhang2025hidiffusion} modify network structures but risk suboptimal performance.
SVG~\citep{xi2025sparse} explores sparsity within the attention module but achieves only limited acceleration.
Jenga~\citep{zhang2025training} employs dynamic token carving to reduce redundancy in attention computation, but its generated content deviates from the original model.

\noindent \textbf{Efficient video generation.}  
The main challenge of efficient video generation lies in the quadratic complexity of attention mechanisms~\citep{cai2025ditctrl}, especially under high-resolution settings. 
Several works~\citep{cai2023efficientvit, xie2024sana, wang2020linformer, choromanski2020rethinking, yu2022metaformer, katharopoulos2020transformers} transform attention into linear operations to reduce computational cost,
while others adopt hybrid strategies combining local and global attention to select informative token pairs~\citep{xi2025sparse, zhang2025spargeattn, zhang2025fast, xia2025training,ding2025alchemist}.
FlashAttention~\citep{dao2022flashattention} introduces a patch-divided acceleration method, and more recently, SEED-VR~\citep{wang2025seedvr} proposes variable-sized windows near spatiotemporal boundaries to better support long video sequences.
For faster sampling, rectified flow models with straight ODE trajectories~\citep{esser2024scaling} have been proposed for text-to-image (T2I) tasks. 
%
%
However, applications in text-to-video (T2V) remain limited due to the added complexity of temporal dynamics. 
While some methods~\citep{ding2024dollar, zhang2025flashvideo} reduce denoising steps, they are still constrained to long-frame, high-resolution settings and often overlook resolution-domain inconsistency.
\citet{tian2025training,yang2025rectifiedhr} further explored stage-wise denoising, but their framework is restricted to a single model and only supports two-resolution transfer, without investigating how different step divisions affect generation quality, thereby offering limited acceleration benefits.
These observations motivate our design of a resolution-dynamic T2V generation framework.

%% file: sections/3_method.tex
\section{Method}\label{sec:method}%
\begin{figure*}[t]
\centering 
\includegraphics[width=0.98\linewidth]{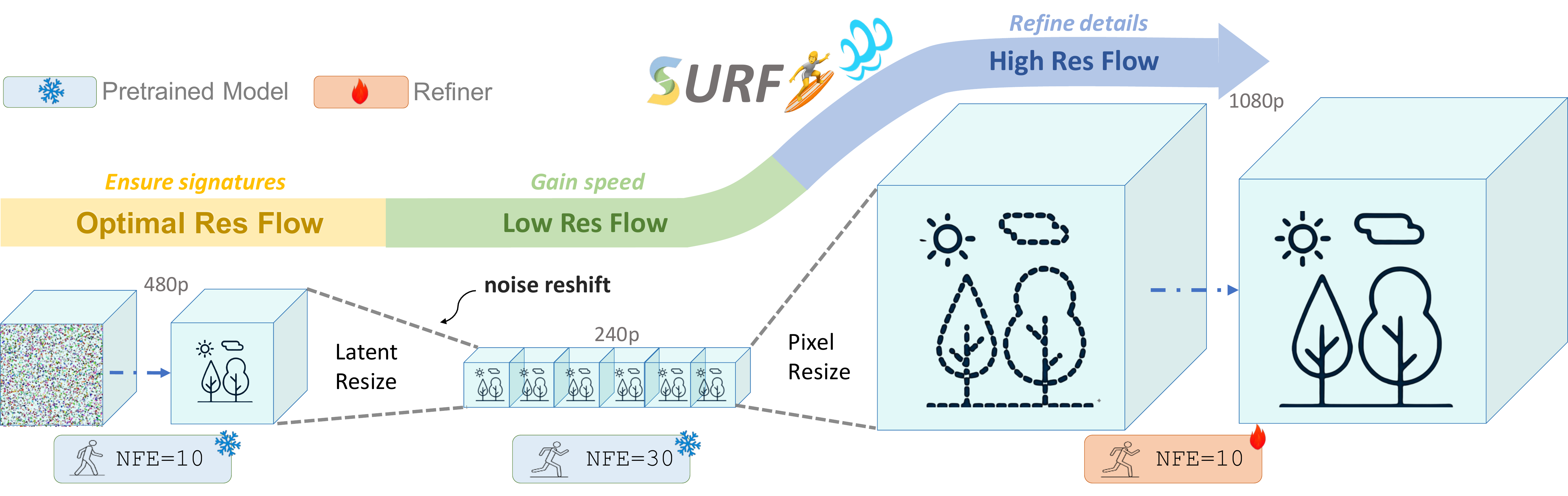} 
\vspace{-10pt}
\caption{
\textbf{Overview of the \method.}
Our framework employs a powerful pretrained model (e.g., Wan~2.1) and a lightweight Refiner network.
Together, they execute the denoising process through an OptimRes-LowRes-HighRes flow~(\method), producing outputs with signature~(\textit{e.g.,} layout, semantic, motion) closely match those of the base model (Wan~2.1), while achieving huge speed up.
}
\label{fig:main-pipeline}
\end{figure*}
In this work, we propose \method, a framework that significantly improves the efficiency of pretrained video generation models while maximally preserving their generative \prior.

\subsection{Overall Pipeline.}
The overall framework of \method is shown in Fig.~\ref{fig:main-pipeline}.
Current video acceleration methods often operate at a fixed latent scale. While strategies~\citep{zhang2025spargeattn,xi2025sparse,yang2025sparse} that exploit the sparsity of video tokens within attention modules can reduce computational costs, their acceleration potential remains limited. 
Moreover, adopting an aggressive token dropping policy—even retaining specifically selected important tokens—inevitably degrades the learned generative \prior.
Therefore, we analyze that the key factors that influence the generation speed are the \textbf{resolution} and the \textbf{number of denoising steps}.

We design the pipeline under this principle: we infer at optimal resolution and downsample the latent to generate low-resolution previews; then we add more details to the preview with fewer denoising steps at target high-resolution. 
\method introduces a dynamic scaling mechanism that allocates a variable number of tokens according to the denoising timestep.
Instead of permanently discarding tokens, we resize the latent scale to modulate the token count, thereby ensuring that the global information from the entire token set is preserved.

Specifically, in the preview stage, we retain the full capacity of a powerful pretrained model to establish the global structure with content and accelerate the whole process by dynamic scaling mechanism;
In the refine stage, we switch to a lightweight model to both accelerate the process and enrich the details. 
\subsection{Preview stage}
In this stage, we aim to leverage the strong generative capabilities of a given pretrained model for generating a low-resolution preview. 
We expect the low-resolution preview to preserve the \prior~(layout, semantic, motion) of the given pretrained model.
A straightforward solution is to let the pretrained model infer on the lower resolution. However, we find that each model has its own optimal resolution; inferring on mismatched resolution causes severe degradation.
Inspired by the fact that the overall structure is determined by the initial denoising steps, we introduce a progressively downsampling method. 

\noindent \textbf{Noise reshifting.}  During the denoising phase, we begin with the initial gaussian noise $z_1 \in R^{b \times c \times f \times h \times w}$ and progressively downsample the clean latent representation $z_0$ with reduced spatial dimensions $h’ \times w’$. 
Here $b$, $c$, $f$ and $h, w$ denote the batch size, number of channels, number of frames, and resolution of the model's optimal generation, respectively.
Our strategy is fundamentally structured around a turning point step k along the denoising trajectory and divides into pre-k steps and post-k steps. %

Before reaching k(pre-k steps), the latent representation is denoised through an ODE-based flow matching approach, described by the following iterative update:
\begin{equation}
\mathbf{z}_0 = \mathbf{z}_1 + \int_{1}^{0} \mathbf{u}_\theta(\mathbf{z}_t, t) \, dt
\label{eq:flow_matching}
\end{equation}
where $\mathbf{u}_\theta(\mathbf{z}, i)$ represents the model-predicted direction function. 
Upon reaching step $k$, we estimate the clean latent representation as
$\hat{\mathbf{z}}_0 = \mathbf{z}_k - \sigma_k \cdot \mathbf{u}_\theta(\mathbf{z}_k, k)$,
where $\sigma_k$ denotes the noise standard deviation at step $k$.
Subsequently, we apply a spatial downscaling operation to the estimated latent,
$\hat{\mathbf{z}}_0^\downarrow = \text{Downscale}(\hat{\mathbf{z}}_0)$.
where the simple linear \text{downscale} operation is implemented in the latent space. This choice is crucial as it effectively reduces spatial resolution while vigilantly preserving essential spatiotemporal coherence.
Afterwards, we reinject noise which is shifted to timestep $k$, into the reduced-resolution latent space, allowing the stochastic denoising process to seamlessly resume.
The reshifting process is represented by:
\begin{equation}
\mathbf{z}_{k-1} = \hat{\mathbf{z}}_0^\downarrow + \sigma_k \cdot \tilde{\bm{\epsilon}}, \quad \tilde{\bm{\epsilon}} \sim \mathcal{N}(\mathbf{0}, \mathbf{I}).
\end{equation}
In post-k steps, latents denoises in lower-resolution to gain speed.
This multi-scale denoising framework enables the model to first gain global semantics at coarser scales, and then get the preview at finer scales.
\subsection{Refine stage}
To alleviate the computational overhead, we employ a lightweight model with 1B parameters, which reduces the time cost per step (architectures and training details can be found in \supp).

\begin{figure}[t]
  \centering
  \includegraphics[width=\linewidth]{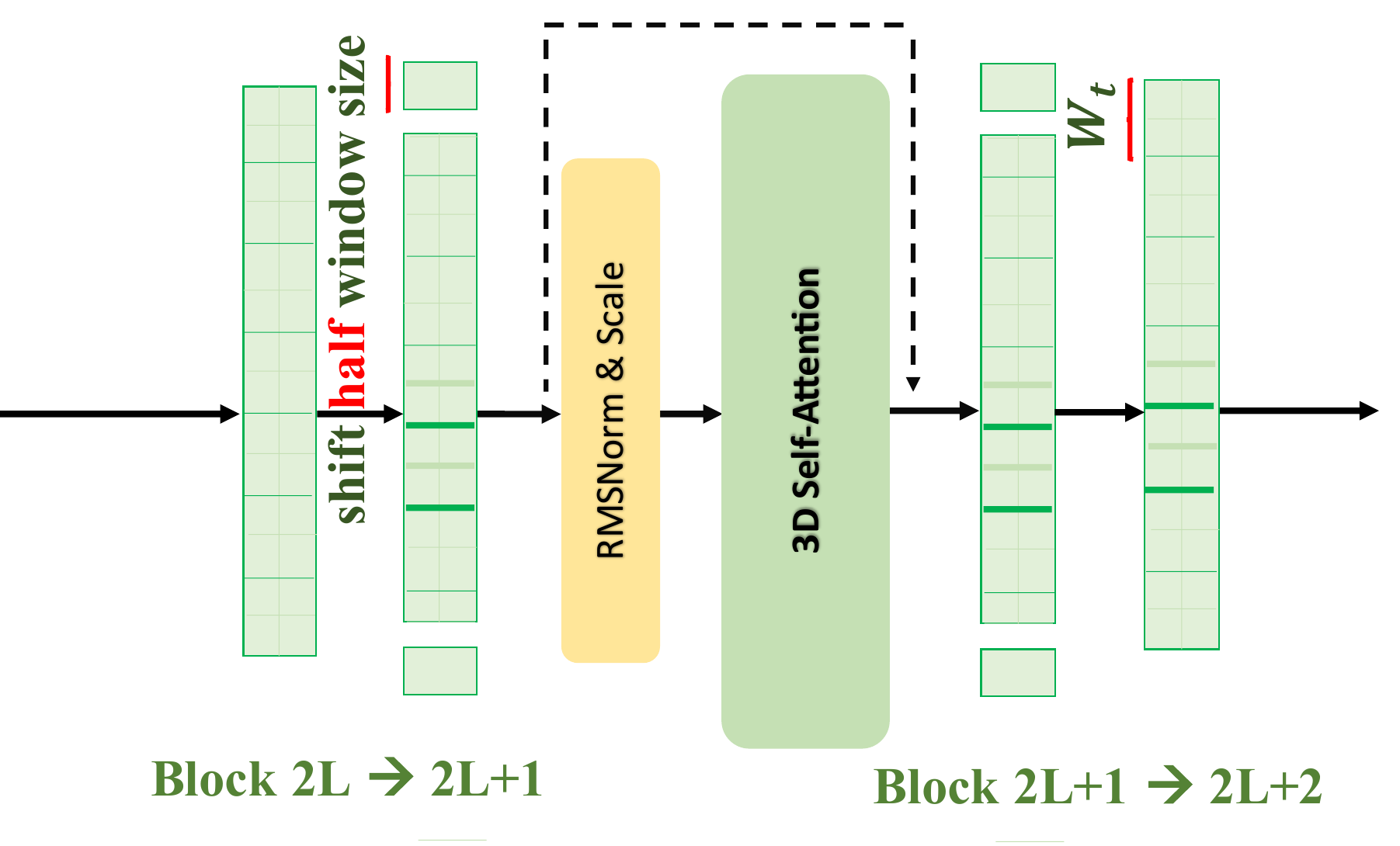}
  \vspace{-25pt}
  \caption{%
    \textbf{Shift window} across adjacent blocks.
  }
  \label{fig:shiftwindow}
\end{figure}
\noindent \textbf{Flow mapping.} We start from the preview and linear upsample it to form blurred low-resolution latents. 
Then we establish a mapping between low-resolution latents $z_{lr}$  and high-resolution latents $z_{hr}$ by modifying the flow-matching~\eqref{eq:flow_matching}, substituting $z_1$ with $z_{lr}$ and $z_0$ with $z_{hr}$. 
Our light-weight Refiner learns $z_{lr}$'s directional information, facilitating high-quality video generation in a few denoising steps. 

\noindent \textbf{Training paradigm.} We first obtain large amounts of high-quality videos and apply both pixel- and latent-level degradations to simulate low-quality video values. Pixel-level degradation simulates blur, while latent-level degradation prevents the task from becoming a trivial super-resolution problem, encouraging the model to exploit its generative ability. These low-quality latents are then paired with the original high-quality latents to form the training data. 
However, with large resolution and long frame numbers, it presents a computational challenge and results in considerable time consumption per denoising step. 

\noindent \textbf{Model structure.}  Given a video feature $\mathbf{X} \in \mathbb{R}^{T \times H \times W}$, our transformer design addresses this challenge by balancing efficiency and temporal connectivity. The first transformer block applies \textit{regular window attention} with a $t \times H \times W$ window. Specifically, $\mathbf{X}$ is divided into $\left(\frac{T}{t}+1\right) \times H \times W$ windows, where each window spans $t$ consecutive frames along the temporal dimension while fully covering the spatial dimensions. This windowed partition allows local temporal interactions to be captured before expanding to broader temporal receptive fields in deeper blocks. 
Our solution further integrates a cyclic shift-window strategy into the 3D self-attention modules shown in Fig.~\ref{fig:shiftwindow}. 
This strategy, embedded within Transformer blocks, establishes full temporal connectivity through a two-phase cycle across layers. 
Each consecutive layer pair collaboratively connects all frames while maintaining computational efficiency. 

In detail, Transformer block $2L$ applies 3D self-attention to non-overlapping temporal windows of size $W_t$, adjusting position embeddings for local awareness. 
The subsequent Transformer block $2L+1$ applies a temporal shift of $S_t=\tfrac{W_t}{2}$ to the input feature, then partitions it into windows of size $W_t$ for attention computation. 
The cyclic-shifted attention mechanism can be expressed as: 
\begin{equation}
\begin{aligned}
\mathbf{X}^{(2L)} &= \text{Attention3D}(\text{Partition}(\mathbf{X}, W_t)), \\
\mathbf{X}_{\text{shifted}} &= \text{CyclicShift}\!\left(\mathbf{X}^{(2L)}, \tfrac{W_t}{2}\right), \\
\mathbf{X}^{(2L+1)} &= \text{Attention3D}\!\left(\text{Partition}(\mathbf{X}_{\text{shifted}}, W_t), \text{Mask}\right).
\end{aligned}
\end{equation}
When we shift the window by half the window size forward, one boundary window (e.g., the first window) contains temporally unrelated halves. Therefore, we apply an attention mask to separate them. 
We also employ position-frequency embeddings with 3D RoPE~\citep{su2024roformer} within each window to avoid the resolution bias introduced by fixed positional embeddings. 
This two-block cycle (unshifted/shifted layers) guarantees global temporal connectivity, significantly reducing memory and accelerating attention calculation for large latent tensors.

%% file: sections/4_exp.tex
\section{Experiments}\label{sec:exp}
%

\begin{table*}[t]
    \centering
    \small
    \caption{
        \textbf{Quantitative comparison on Wan~2.1.} We report evaluations of the baseline~(row 1), step and attention optimization methods~(row 2-5) and \method~(row 6).
        NFE=50.
        The highest score is in \textbf{bold} and the second highest is \underline{underlined}.
        Abbreviations: QS (Quality Score), AQ (Aesthetic Quality), DD (Dynamic Degree), 
        MS (Motion Smoothness), OC (Overall Consistency), SA (Semantic Adherence), PC (Physics Commonsense).
    }
    \label{tab:vbench_selected}
    \renewcommand{\arraystretch}{1.2}
    \setlength{\tabcolsep}{8.5pt}{
    \begin{tabular}{l|ccccc|cc|ccc}
        \toprule
        \multirow{2}{*}{\textbf{Method}} & \multicolumn{5}{c|}{\textbf{General Scene}} & \multicolumn{2}{c|}{\textbf{Physical Scene}} & \multicolumn{3}{c}{\textbf{Computation Loads}} \\
        \cline{2-6} \cline{7-8} \cline{9-11}
         & QS$_{\textcolor{purple}{\uparrow}}$ 
        & AQ$_{\textcolor{purple}{\uparrow}}$ 
        & DD$_{\textcolor{purple}{\uparrow}}$ 
        & MS$_{\textcolor{purple}{\uparrow}}$ 
        & OC$_{\textcolor{purple}{\uparrow}}$ 
        & SA$_{\textcolor{purple}{\uparrow}}$ 
        & PC$_{\textcolor{purple}{\uparrow}}$ 
        & Time$_{\textcolor{purple}{\downarrow}}$ 
        & Speed$_{\textcolor{purple}{\uparrow}}$ 
        & PFLOPs$_{\textcolor{purple}{\downarrow}}$ \\
        \hline
        Wan~2.1~\citep{wan2025} & \underline{83.31} & \textbf{66.9} & 63.89 & 97.65 & 27.08 & \textbf{41.82} & \textbf{45.45} & 3497 (58min) & 1$\times$  & 658.5 \\
        \hline
        30\%step~\citep{wan2025} & 77.92 & 58.43 & 56.94 & 96.95 & 24.56 & 18.18 & 16.36 & 1049 & 3.34$\times$  & 197.5 \\
        50\%step~\citep{wan2025} & 81.51 & 63.52 & 66.67 & 96.99 & 25.90 & 25.45 & 23.64 & 1748 & 2.00$\times$  & 329.2 \\
        SVG~\citep{xi2025sparse} & \textbf{83.36} & 65.6 & \underline{68.06} & 97.69 & \underline{27.32} & 25.45 & 20.00 & 2712 & 1.29$\times$  & 429.9 \\
        DMD~\citep{fastvideo2024} & \underline{83.31} & 66.11 & 52.78 & \textbf{98.96} & 26.77 & \underline{34.55} & 30.91 & \underline{282} & \underline{12.40}$\times$  & \underline{39.5} \\
        \hline
        Ours & 83.26 & \underline{66.86} & \textbf{72.22} & \underline{97.95} & \textbf{27.38} & \textbf{41.82} & \underline{38.18} & \textbf{278} & \textbf{12.58}$\times$  & \textbf{34.3}\\
        \bottomrule
    \end{tabular}
    }
\end{table*}
\begin{figure*}[t]
\centering 
\includegraphics[width=0.98\linewidth]{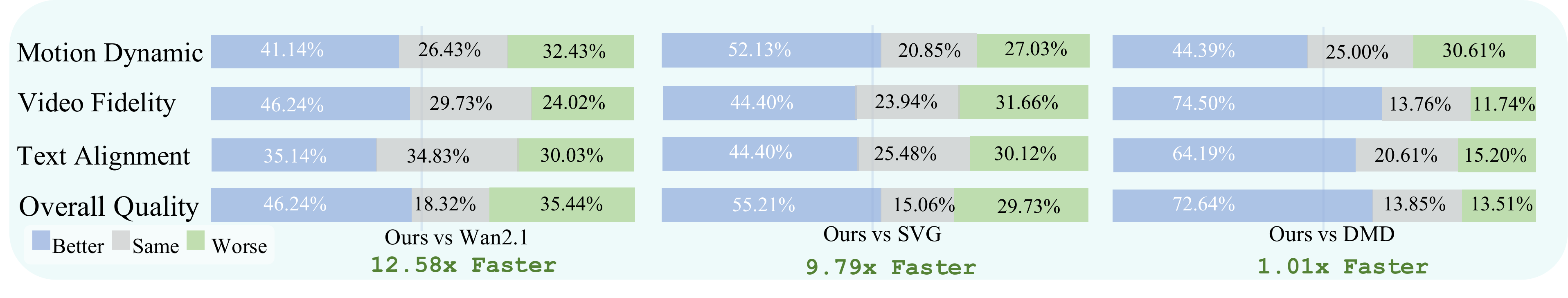} 
\vspace{-10pt}
\caption{\textbf{User study.} We report pair-wise preference rates. \method achieves comparable quality to Wan~2.1 with huge speed up.}
\label{fig:user_study}
\vspace{-8pt}
\end{figure*}
\subsection{Implementation Details}
\noindent \textbf{Settings.} The Refiner of \method is trained on 24 NVIDIA A800 GPUs (80 GB each) with a total batch size of 24. 
We finetune the transformer using the AdamW optimizer with a learning rate of 5e-5.
We create a synthetic dataset of 100k LR-HR video frame pairs following the methodology in \citet{wang2021realesrgan}.
To optimize training efficiency and stabilize convergence, we employ a progressive training strategy where the model is trained at incrementally increasing frame num.
We first train \method on 21-frame inputs for 1k iterations, and then extend the input length to 81 frames and finetune the model for 4k iterations.

\noindent \textbf{Evaluation and metrics.} For speed assessment, we report the DiT forward pass time on NVIDIA A800, as the VAE decoding component remains constant across all configurations. 
We also report FLOPs to provide an intuitive comparison of computational complexity. 
For qualitative evaluation, we construct a video dataset of 381 low-resolution videos with prompts from VBench~\citep{huang2024vbench}, Videophy~\citep{bansal2024videophy} and PhyGenBench~\citep{meng2024towards}; 
We evaluate each prompt with a fixed seed to ensure reproducibility. 
Additionally, we conducted a user study to assess human preference rates between \method and various efficient generation methods.
As for 1080p video generation, most existing acceleration methods are not capable of handling this setting. 
Therefore, for a fair comparison, we benchmark our approach against super-resolution(SR) methods and adopt several widely used video SR assessment metrics. 
Specifically, we employ DINO~\citep{caron2021emerging} and CLIP~\citep{radford2021learning} to evaluate frame quality via feature similarity across frames;
LAION aesthetic predictor~\citep{schuhmann2022laion} to assess artistic and beauty value perceived by humans towards each video frame, and DOVER~\citep{wu2023exploring} to measure overall video quality.

\subsection{Comprehensive Evaluations}
We evaluate our method from both efficiency and quality perspectives.
Specifically, We compare it with the Wan~2.1 baseline, which employs the FlowMatch scheduler with 50 NFE~(number-of-evaluations), as well as accelerated Wan~2.1 variants utilizing 30\% and 50\% of the original steps. 
Additionally, we compare our approach with two commonly adopted acceleration techniques: the sparse attention mechanism used in SVG~\citep{xi2025sparse} and the bidirectional video distribution matching distillation~(DMD) method~(implemented following \citet{fastvideo2024}).

\noindent \textbf{Efficiency analysis.}
The quadratic cost of attention makes token count the key bottleneck for efficiency.
\method addresses this by adapting latent scales across denoising stages.
The first stage processes inputs at a resolution equivalent to 480p, the second at 240p, and the third utilizes a smaller model with approximately 5$\times$ fewer hidden dimensions and 2.5$\times$ fewer attention heads, resulting in negligible FLOPs compared to the original model.
As shown in Tab.~\ref{tab:vbench_selected}, our method achieves performance comparable to Wan~2.1 at 720p resolution while reducing inference time by 12×. 
Furthermore, for 1080p video generation, our approach achieves a remarkable 43$\times$ acceleration compared to a direct application of Wan~2.1 at the same resolution.
In contrast, other methods such as SVG operate on latents of the same scale, achieving acceleration primarily through hardware-efficient tensor layouts, which offer limited reduction in redundant tokens. 
DMD, on the other hand, focuses on reducing the number of inference steps. 
\method distinguishes itself by leveraging the intrinsic properties of the denoising process, adapting latent representations at different scales to different stages of denoising. 
This strategy achieves substantial acceleration while preserving crucial \prior information. 
Although the overall runtime of distillation-based methods is comparable to ours, they fail to preserve signatures of original models and introduce quality degradation~(\textit{e.g.}, misaligned limbs) in generated videos.

\noindent \textbf{Quality analysis.} We evaluated our model’s generative performance from two perspectives: general video metrics and physics-focused assessments (as original Wan~2.1 achieves best in physical plausibility so we want to prove \method has preserved this characteristics). 

\begin{figure*}[t]
\centering 
\includegraphics[width=1\linewidth]{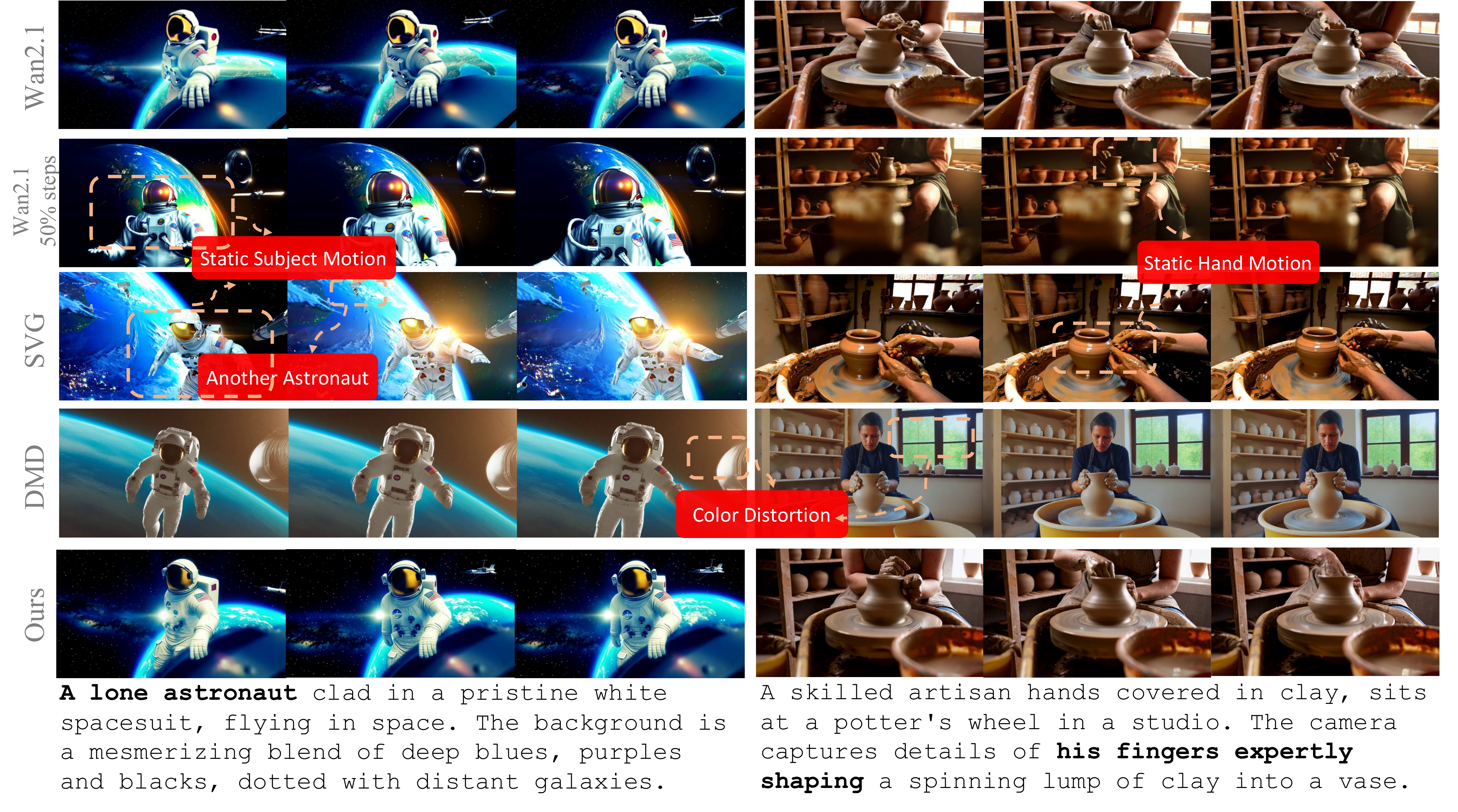} 
\vspace{-22pt}
\caption{%
\textbf{Qualitative comparisons.} Our method achieves up to 12× speedup while maintaining \prior of base model. Unreasonable contents are marked in \textcolor{orange}{orange}. Rather than aimless camera panning, \method generates high fidelity videos with semantically aligned motion. }
\label{fig:Compare}
\end{figure*}

As illustrated in Tab.~\ref{tab:vbench_selected}, our generated video exhibits visual quality comparable to Wan~2.1, while slightly surpassing SVG and outperforming the DMD method.
For further validation, we provide qualitative comparisons in Fig.~\ref{fig:Compare}.
We also conducted a perceptual evaluation using a standard win-rate methodology. We have designed questionnaire with 24 randomly selected videos from the aforementioned test datasets. A total of 37 researchers in the field of video generation were asked to evaluate the results along four dimensions: Motion, Fidelity, Semantics, and Overall Quality.
\begin{table}[t]
\centering
\caption{\textbf{Plug-in integration} with other acceleration methods and different model architectures.}
\vspace{-10pt}
\label{tab:adaptation}
\resizebox{\linewidth}{!}{%
\begin{tabular}{l|cccc}
\toprule
\textbf{Method} & \textbf{SA}$_{\textcolor{purple}{\uparrow}}$ & \textbf{PC}$_{\textcolor{purple}{\uparrow}}$ & \textbf{NFE/Time}$_{\textcolor{purple}{\downarrow}}$ & \textbf{Speed}$_{\textcolor{purple}{\uparrow}}$  \\
\hline
HunyuanVideo~\citep{kong2024hunyuanvideo}              & 29.09 & 27.27 & 50/3081 & 1$\times$  \\ %
+SparseAttn~\citep{zhang2025training}               & 30.91 & 23.64 & 50/2775 & 1.1$\times$  \\ %
+ \method                 & 43.64 & 45.45 & 50/356 & 8.7$\times$  \\
\hline
AccVideo~\citep{zhang2025accvideo}                  & 32.73 & 23.64 & 5/340 & 1$\times$  \\
+ \method                 & 36.36 & 38.18 & 5/265 & 1.3$\times$   \\
\bottomrule
\end{tabular}
}
\vspace{-18pt}
\end{table}
The outcomes, summarized in Fig.~\ref{fig:user_study}, demonstrate that our method achieves performance comparable to the original model while surpassing many existing acceleration approaches from a human perspective. 
From extensive examples, we observe that SVG suffers from limited robustness where main subjects often remain static and adds unreasonable details;
In contrast, DMD tends to introduce unnatural color artifacts and produces grainy videos with reduced fidelity.
Moreover, our method can be naturally extended to support 1080p video generation.
Since most baseline models do not natively support this resolution, we evaluate 1080p results from the perspective of video super-resolution (SR).
Specifically, we randomly select 100 samples from the aforementioned test datasets and conduct comparisons with existing SR approaches, aiming to assess the effectiveness of our method in generating high-resolution videos with enhanced detail and fidelity.
%
%
%
The detailed results, as shown in Tab.~\ref{tab:vsr_compare}, indicate our method's strong performance. Notably, although GAN-based RealBasicVSR achieves competitive scores on some metrics, its outputs frequently exhibit excessive smoothing, which does not align with human perceptual preferences;
Diffusion-based VEnhancer also demonstrates strong generative capabilities, however, its outputs often undergo significant deviations from the input, contradicting the principle of enhancing visual quality while preserving fidelity.
%

\noindent \textbf{Compatibility analysis.} Furthermore, \method can function as a plug-in compatible with various diffusion model architectures and acceleration techniques.
When combined with sparse attention~\cite{zhang2025training}, it achieves a 8.7× speedup on HunyuanVideo~\citep{kong2024hunyuanvideo}.
Additionally, it can be adapted to step-distillation models AccVideo~\citep{zhang2025accvideo} and achieves a 1.3× speedup.
Detailed results are presented in Tab.~\ref{tab:adaptation}.

\begin{figure*}[t]
    \begin{center}
    \includegraphics[width=\linewidth]{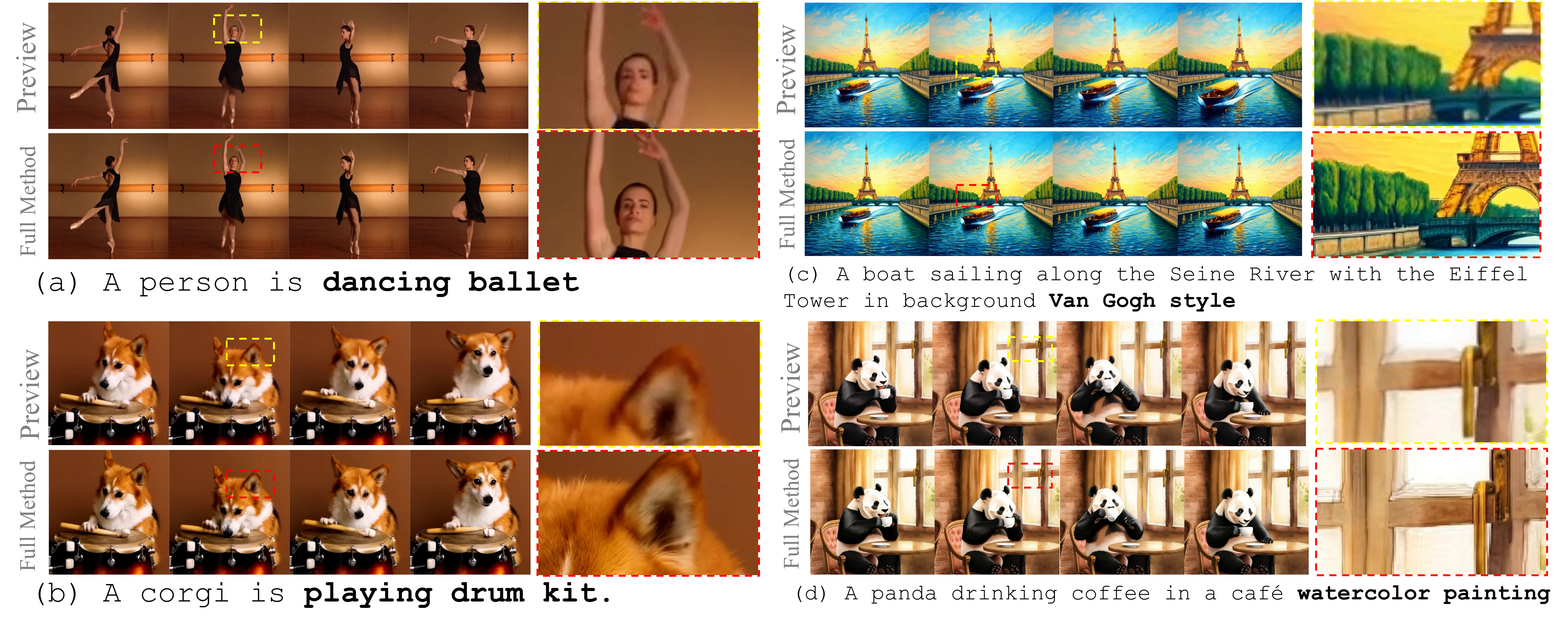}
    \end{center}
    \vspace{-22pt}
    \caption{\textbf{Generation results for each stage}. We mark regions with artifacts and lacking detail in the preview videos using \textcolor{yellow}{yellow} boxes, while improvements from the Refiner are highlighted in \textcolor{red}{red}. Zoom in for a better view. 
    } 
    \label{fig:stage2_demo}
    \vspace{-10pt}
\end{figure*}

\begin{table*}[t]
    \centering
    \begin{minipage}{0.48\textwidth}
        \centering
        \caption{\textbf{Quantitative comparison} with video super-resolution method~(1080p).}
        \vspace{-5pt}
        \scriptsize
        \setlength{\tabcolsep}{4.5pt}
        \renewcommand{\arraystretch}{1.1}
        \begin{tabular}{lcccc|c}
        \toprule
        \textbf{Method} & \textbf{DINO}$_{\textcolor{purple}{\uparrow}}$ & \textbf{CLIP}$_{\textcolor{purple}{\uparrow}}$ & \textbf{LAION}$_{\textcolor{purple}{\uparrow}}$ & \textbf{DOVER}$_{\textcolor{purple}{\uparrow}}$ & \textbf{NFE/Time}$_{\textcolor{purple}{\downarrow}}$ \\
        \hline
        RealBasicVSR & 93.40 & 94.83 & 61.07 & \underline{80.25} & 1/162.1s \\
        Upscale-a-Video   & 93.47 & 95.71 & 60.93 & 71.40 & 30/2517.7s \\
        VEnhancer    & 93.55 & 96.02 & \underline{63.46} & 79.78 & 15/2467.6s \\
        STAR  & \underline{93.68} & \textbf{96.59} & 60.81 & 63.64 & 14/912.7s \\
        Ours & \textbf{93.75} & \underline{96.30} & \textbf{63.50} & \textbf{81.20} & 10/76.5s \\
        \bottomrule
        \end{tabular}
        \label{tab:vsr_compare}
    \end{minipage}
    \hfill
    \begin{minipage}{0.48\textwidth}
        \centering
        \caption{\textbf{Ablation study} of step division in preview stage. The recommended setting is in gray.}
        \vspace{-5pt}
        \scriptsize
        \setlength{\tabcolsep}{8pt}
        \renewcommand{\arraystretch}{1.1}
        \begin{tabular}{lccccc|c}
        \toprule
        \textbf{Setting} & \textbf{QS}$_{\textcolor{purple}{\uparrow}}$ & \textbf{AQ}$_{\textcolor{purple}{\uparrow}}$ & \textbf{DD}$_{\textcolor{purple}{\uparrow}}$ & \textbf{MS}$_{\textcolor{purple}{\uparrow}}$ & \textbf{OC}$_{\textcolor{purple}{\uparrow}}$ & \textbf{Time}$_{\textcolor{purple}{\downarrow}}$ \\
        \hline
        \textbf{5-35}  & 82.21 & \underline{63.45} & 66.67 & \textbf{98.23} & 27.12 &  201s \\
        \rowcolor{gray!40}
        \textbf{10-30} & 82.01 & 62.87 & \textbf{70.83} & \underline{98.16} & 27.51 & 252s \\
        \textbf{20-20} & 81.19 & 62.54 & \underline{69.44} & 98.05 & \textbf{27.57} & 369s \\
        \textbf{30-10} & 80.78 & 61.37 & \textbf{70.83} & 98.05 & \underline{27.52} &  481s \\
        \textbf{40-0}  & \textbf{82.89} & \textbf{66.57} & 68.06 & 97.70 & 27.35 & 610s \\
        \bottomrule
        \end{tabular}
        \label{tab:step_config}
    \end{minipage}
    \vspace{-12pt}
\end{table*} 
\subsection{Stage-wise Visualization}
This section presents visualizations of \method, providing an intuitive comparison between the preview and the final results in overall layout similarity and refinement capability.
As shown in top row of Fig.~\ref{fig:stage2_demo}, our design first prioritizes an efficient retention of the inherent signatures of the powerful model to ensure robust structural layout, semantic alignment, and motion dynamics. 
For instance, the ballet dancer's movements are demonstrably smooth and natural.
Furthermore, our method is adept at style-aware video synthesis, exemplified by results such as a Van Gogh-style tower and a watercolor panda.
Then the well-trained Refiner refines intricate details and mitigates generation artifacts. 
In second row, \method further enhances visual fidelity of existed videos, including clearer facial and hand structures in (a), refined fur on the corgi in (b), enhanced texture and structural details in (c), and more semantically consistent structural and color corrections in (d).
More examples in Fig.~\ref{fig_1} further demonstrate the visual aesthetic and \method-retained of the generated videos.


\begin{table}[t]
    \centering
    \caption{\textbf{Ablation study} of denoising steps in refine stage. 
    We report \method in 720p~(left block) and 1080p~(right block). The recommended setting is in gray.}
    \vspace{-5pt}
    \scriptsize
    \setlength{\tabcolsep}{1.5pt} 
    \renewcommand{\arraystretch}{1.1} 
    \begin{tabular}{c|cccccc|cccc}
    \toprule
    \multirow{2}{*}{\textbf{Step}} & \multicolumn{6}{c|}{\textbf{720p}} & \multicolumn{4}{c}{\textbf{1080p}} \\
    \cline{2-7} \cline{8-11}
     & QS$_{\textcolor{purple}{\uparrow}}$ & AQ$_{\textcolor{purple}{\uparrow}}$ & DD$_{\textcolor{purple}{\uparrow}}$ & MS$_{\textcolor{purple}{\uparrow}}$ & OC$_{\textcolor{purple}{\uparrow}}$ & Time$_{\textcolor{purple}{\downarrow}}$ & DINO$_{\textcolor{purple}{\uparrow}}$ & CLIP$_{\textcolor{purple}{\uparrow}}$ & LAION$_{\textcolor{purple}{\uparrow}}$ & DOVER$_{\textcolor{purple}{\uparrow}}$ \\
    \midrule
    8  & 83.24 &66.90 & 72.22 & 97.94 & 27.34 & 244.5  & 93.70 & 96.28 & 63.48 & 80.52 \\
    9  & 83.17 &66.90 & 70.83 & 97.95 & 27.38 & 247.0  & 93.75 & 96.31 & 63.48 & 80.89 \\
    \rowcolor{gray!40}
    10 & 83.26 &66.86 & 72.22 & 97.95 & 27.38 & 249.0  & 93.75 & 96.30 & 63.54 & 81.20 \\
    11 & 83.22 &67.03 & 70.83 & 97.95 & 27.38 & 251.8  & 93.71 & 96.32 & 63.60 & 81.57 \\
    12 & 83.31 &66.69 & 72.22 & 97.97 & 27.37 & 254.2  & 93.73 & 96.27 & 63.49 & 81.43 \\
    \bottomrule
    \end{tabular}
    \vspace{-10pt}
    \label{tab:denoise_compare}
\end{table}

\subsection{Ablation}
%
In this section, we present ablation studies to investigate the impact of each key component on the trade-off between computational efficiency and generated video quality.

\noindent \textbf{Step division.}
We look into how to balance efficiency and quality of generating a useful preview.
We have found that the division of step range~(i.e., the choice of k) is very important.
Since the efficiency of our method is directly proportional to the number of denoising steps performed at the initial resolution, identifying the optimal step range for the transition is crucial for balancing quality and speed.
With the results presented in Tab.~\ref{tab:step_config}, it reveals a distinct trade-off. Employing an early transition (e.g., after step 5) accelerates the process but leads to a degradation in both layout integrity and motion quality. 
Conversely, delaying the transition (e.g., after step 35) not only increases the inference time but can also compromise the final layout, as the late-stage resolution shift may disrupt an already well-defined structure.
Besides, We observe the denoising process of video generation that the overall compositional layout of the generated content stabilizes after approximately step 10.
Based on this analysis, we identify the 10-30 step range as the optimal configuration. This range effectively preserves the structural and motion quality of the generated video.

\begin{figure}[t]
    \centering
    \includegraphics[width=0.95\linewidth]{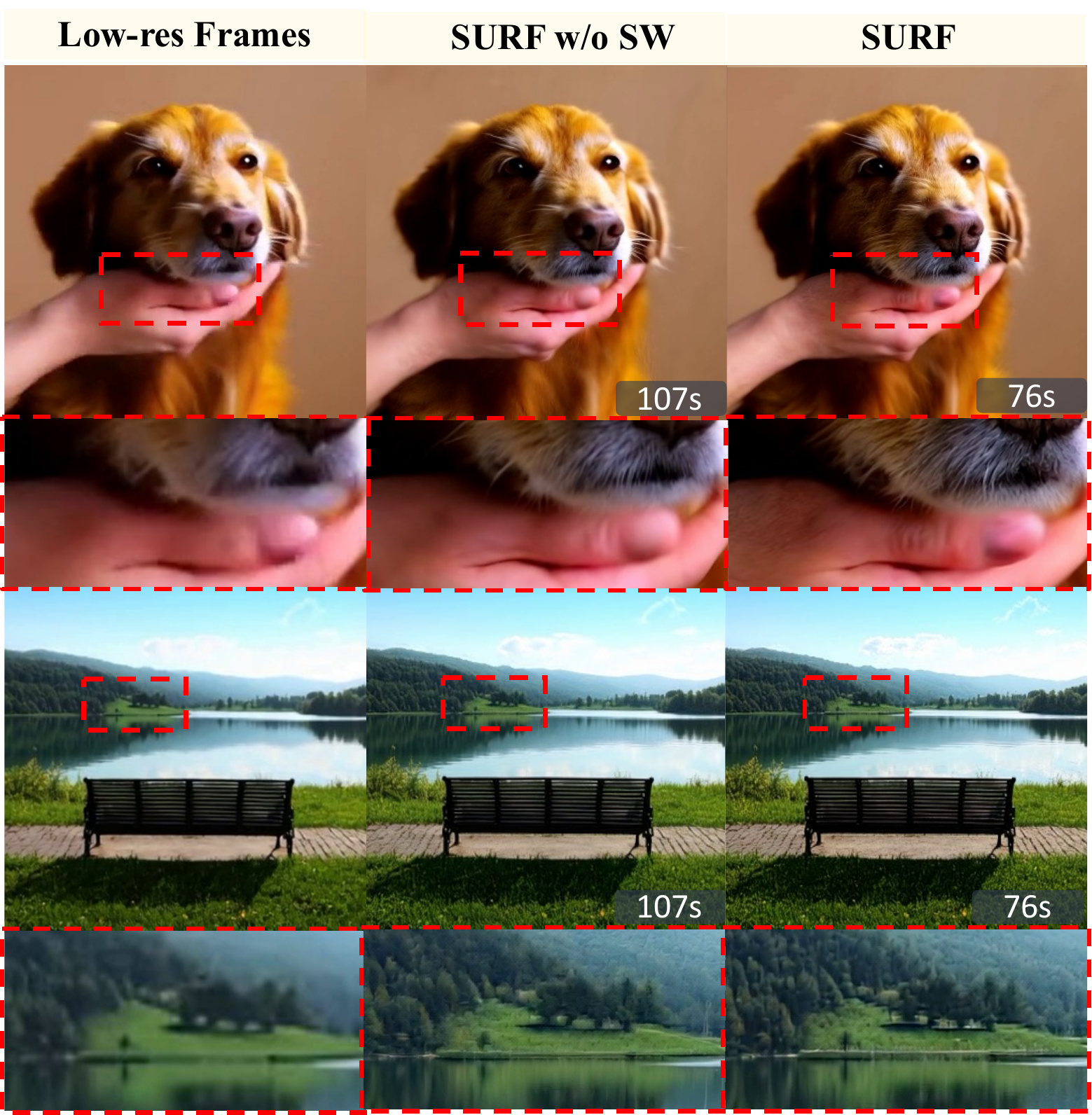}
    \vspace{-6pt}
    \caption{%
    \textbf{Qualitative results for ablation.} 
    SW denotes shift-window attention. The Refiner runtime on 1080p video is shown in the bottom-right corner. 
    Our method achieves shorter runtime while preserving finer details.}
    \label{fig:ablation_shiftwindow}
     \vspace{-15pt}
\end{figure}

\noindent \textbf{Shift-window attention.}
We further investigate the impact of incorporating shift-window attention in Refiner on video generation quality.
As shown in Fig.~\ref{fig:ablation_shiftwindow}, we observe that previously unclear and distorted hands now exhibit clear fingernail contours and appropriate wrinkles, regardless of whether the shift-window mechanism is applied. Additionally, the clarity of distant trees is noticeably enhanced.
our findings suggest that a full receptive field from global attention is not critical for the refinement stage. The visual differences are negligible, indicating that local context modeling is sufficient for enhancing details at this stage. 

\noindent \textbf{Inference hyperparameters.}
We further investigate the influence of common inference-time hyperparameters on video generation performance.
The performance trends across different steps in Refiner are visualized, with corresponding quantitative results summarized in Tab.~\ref{tab:denoise_compare}.
We analyze the effects of varying the number of diffusion steps and highlight the best-performing configurations in \colorbox{gray!40}{gray}.
%
%

%% file: sections/5_conclusion.tex
\section{Conclusion}\label{sec:conclusion}
We present SURF, a simple yet effective framework for efficiently generating high-resolution videos while retaining signatures (\eg, layout, semantics, and motion) of a given pretrained model. SURF delivers substantial improvements in both human perceptual preference and quantitative performance metrics. It can also act as plug-in with other SOTA models. In particular, it attains a 12.5× speedup for generating 5-second, 16-fps, 720p videos with Wan 2.1 and 8.7× speedup for HunyuanVideo.